20 February 2018

# Contactless method to measure 2DEG charge density and band structure in high electron mobility transistor structures

Yury Turkulets and Ilan Shalish*

*Department of Electrical and Computer Engineering, Ben Gurion University of the Negev, Beer Sheva, Israel*

We present a contactless method that is capable of characterizing a high electron mobility transistor heterostructure at the wafer stage, right after its growth, before any production process has been attempted, to provide the equilibrium band structure and the density of charge of the 2-dimensional electron gas in the quantum well. The method can thus evaluate critical transistor parameters and help to screen out low performance wafers before the actual fabrication. To this end, we use a simple optical spectroscopy at room temperature that measures the surface photovoltage band-edge responses in the heterostructure and uses a model that takes into account the effect of the built-in electric fields on optical absorption in the layers and heterojunctions to evaluate bandgaps, band offsets, and built-in fields. The quantum well charge is then calculated from the built-in fields. The main advantage of the method is in its capability to provide information on all the different layers in the typical heterostructure simultaneously in a single measurement. The method is not limited to the high electron mobility transistor structure but may be used on any other heterostructure. It opens the door for a new type of characterization methods suitable for the post-silicon multi-layer multi-semiconductor heterostructure device era.

Heterostructure based transistors, such as high electron mobility transistor (HEMT) and heterostructure bipolar transistor (HBT), requires a careful design of the layer sequence, composition, and thickness, in order to achieve the desired band structures and the specific values of their critical parameters.[1,2] Once the layer sequence has been grown, it has been possible to asses certain band diagram parameters such as bandgaps and band-offsets using photoemission spectroscopies, e.g., x-ray or ultra-violet photoelectron spectroscopies.[3,4,5,6] These spectroscopies are limited to structures of no more than two layers.[7] Spectroscopies based on absorption, overcome this limitation, but nonetheless, face another challenge. The optical transitions they detect are typically red-shifted due to the presence of high built-in electric fields, which renders the band-to-band transition spectra inaccurate for immediate band structure analysis.[8] Nonetheless, it has been shown that surface photovoltage spectroscopy (SPS) could still be used successfully to this end when augmented with numerical simulations.[9,10,11] In most of the practical cases, however, developers of heterostructure devices rely on simulations alone, with their only feedback coming from electrical characterization of fully fabricated devices.[12,13,14] Here, we present a fully empirical method that *does not require simulations*. It also *does not require any fabrication* process and may be carried out on as-grown wafers.

We have recently introduced the channel photocurrent spectroscopy method that provides the desired information not only at equilibrium but also under all operating conditions of the transistor.[15] That method was based on measurement of photocurrent in the HEMT channel under various gate voltages, and therefore required a *full transistor fabrication*. Here, we present a *contactless* method, based on surface photovoltage spectroscopy that provides the desired parameters of the transistor in *equilibrium*.[16] The advantage of the proposed method is in its ability to provide feedback on the structure and its properties right after the growth. Thus, it may be used to screen out low performance wafers saving the cost and the time of a full fabrication process.

Both the theoretical and the experimental parts of this work will be presented for the special case of a plain AlGaN/GaN HEMT structure, without limiting the generality of the method which is suitable for many other combinations and numbers of different semiconductor layers.

Upon absorption of light in an AlGaN/GaN HEMT structure, electron-hole pairs are generated in either the GaN or the AlGaN layers. The photo-generated pairs are separated by the internal electric fields in the layers. The electrons are transferred by the internal electric fields towards the quantum well (QW), increasing the carrier concentration in the well. At the same time, holes are swept away from the heterojunction. The resulting rise in the 2-dimentional electron gas (2-DEG) electron concentration increases the conductivity in the channel. Under a constant drain-to-source voltage this excess charge gives rise to a photocurrent. The same charge separation also modifies the built-in fields resulting in changes to the band bending of both layers, which is observed as a change in the surface photovoltage.

The channel photocurrent may be expressed in terms of the carrier concentration in the 2-DEG:

$$I_{PH}(hv) = V_{DS} \cdot q \cdot \Delta n_S(hv) \cdot \mu \frac{L}{W} \qquad (1)$$

where $V_{DS}$ is the drain-source voltage, $q$ – the electron charge, $\Delta n_S(hv)$ – photo-generated excess sheet charge concentration in the 2-DEG, $\mu$ - electron mobility in the 2-DEG, $L$ and $W$ are the length and width of the 2-DEG channel, respectively.

We have previously modeled the effect of the built in fields on photon absorption and its manifestation in the *channel photo-current*.[15] The same effect may actually be observed in any spectroscopy that is related to absorption of photons in semiconductor. It causes the band-to-band photo-response to commence well below the bandgap energy and increase gradually towards the actual bandgap energy. The generated excess carriers give rise to a photocurrent given by:[15]

$$I_{PH}(hv) = I_{PH}(Eg) exp\left[-\left(\frac{Eg - hv}{\Delta E}\right)^{3/2}\right] \qquad (2)$$

Where $I_{PH}$ is the measured photocurrent normalized to the surface reflectance, i.e., divided by the surface transmission $1-R(hv)$,[17] $I_{PH}(E_g)$ is the photocurrent at the bandgap photon energy following the rise, $R(hv)$ – spectral reflectance from the sample surface, $E_g$ – the bandgap energy of the layer material, $hv$ – the photon energy, and $\Delta E$ is given by:

$$\Delta E = \left[\frac{3qF\hbar}{4\sqrt{2m}}\right]^{2/3} \qquad (3)$$

where F is the built-in field, $\hbar$ – the reduced Planck constant, and $m^*$ is the reduced effective mass in the layer material. Combining Eqs. 1 and 2, we get an expression for the photo-generated excess carrier concentration in the 2-DEG as a function of photon energy:

$$\Delta n_S(hv) = \Delta n_S(Eg) exp\left[-\left(\frac{Eg - hv}{\Delta E}\right)^{3/2}\right] \qquad (4)$$

Photo-generated excess carriers in the 2DEG affect the electric field and, consequently, the band bending in the layer. As the top AlGaN layer holds at the both ends equal charges of opposite signs (ionized surface states vs. 2DEG charge), it may be treated as a parallel-plate capacitor.[18] The GaN layer, on the other hand, is in a state of accumulation at the interface with the AlGaN forming a QW. The effect of electric field on absorption is the strongest at the point where the electric field is the highest. The maximum electric field is reached at the interface. However at that point, electrons cannot be excited to the conduction band minimum, because the first eigenstate is higher than the bottom of the conduction band. The lowest energy point for the first eigenstate is where it meets the bottom of the conduction band, which is therefore the point where absorption starts. Within an infinitesimally small depth around this point the band bending may be assumed linear. Hence, the electric field around this point will be approximately constant. In each of the layers, the electric displacement field may generally be expressed by:

$$\vec{D} = \varepsilon_S \vec{F} + \Sigma \vec{P} \qquad (5)$$

where $\vec{D}$ is the electrical displacement field, $\Sigma \vec{P}$ – the sum of all polarization dipoles (zero for non-polar materials), and $\varepsilon_S$ is the permittivity of the layer material. Rewriting Eq. 5 in terms of the 2DEG sheet charge density we get:

$$n_S + \Delta n_S(hv) = \frac{1}{q}\left(\varepsilon_S \frac{\phi_{BBD} + \Delta \phi_{BB}(hv)}{t} + \Sigma \vec{P}\right) \qquad (6)$$

where $n_S$ – 2-DEG sheet charge density in dark, $\Phi_{BBD}$ –potential across the layer in dark, $\Delta\Phi_{BB}(hv)$ - change in the layer potential upon illumination with photons of energy, $hv$, and $t$ – the layer thickness. Piecewise representation of Eq. 6 yields:

$$\begin{cases} n_s = \frac{1}{q}\left(\varepsilon_s \frac{\phi_{BBD}}{t} + \sum \vec{P}\right) \\ \Delta n_s(hv) = \frac{\varepsilon_s \Delta \phi_{BB}(hv)}{qt} \end{cases} \quad (7)$$

Substituting Eq. 7 into Eq. 4, we obtain an expression for the effect of the built-in field (Franz-Keldysh effect) on the photovoltage:

$$\Delta\phi_{BB}(hv) = \Delta\phi_{BB}(Eg) exp\left[-\left(\frac{Eg-hv}{\Delta E}\right)^{3/2}\right] \quad (8)$$

where $\Delta\Phi_{bb}(E_g)$ is the change in the potential across the layer observed when the photon energy equals the bandgap energy, surface transmission normalized (divided by $1-R(hv)$).

Equation 8 may be transformed into the following linear form:

$$y(hv) = \left[ln\left(\frac{\Delta\phi_{BB}(Eg)}{\Delta\phi_{BB}(hv)}\right)\right]^{2/3} = \frac{Eg-hv}{\Delta E} \quad (9)$$

The slope of the linear function in Eq. 9 may then be used in Eq. 3 to obtain the electric field in the layer. Once the electric fields in both the GaN and the AlGaN layers are known, the 2-DEG charge density is readily obtained from the discontinuity in the electrical displacement field across the interface:[19,20]

$$qn_s = \left(\varepsilon_s \vec{F}_{AlGaN} + \vec{P}_{SP,AlGaN} + \vec{P}_{PE,AlGaN}\right) \\ -\left(\varepsilon_s \vec{F}_{GaN} + \vec{P}_{SP,GaN}\right) \quad (10)$$

where $\vec{F}_{AlGaN}$, $\vec{P}_{SP,AlGaN}$, and $\vec{P}_{PE,AlGaN}$ are electric field, spontaneous, and piezoelectric polarization in the AlGaN, and $\vec{F}_{GaN}$, and $\vec{P}_{SP,GaN}$ are the electric field, and the spontaneous polarization in GaN, respectively.

To test the model experimentally, we used a AlGaN/GaN heterojunction. The heterostructure was grown by metal organic chemical vapor deposition (MOCVD) on c-plane sapphire. The layer sequence was an AlN nucleation layer, 2 μm of undoped GaN, 11 nm of $Al_{30}Ga_{70}N$, and 1.5 nm GaN cap layer. Device fabrication and details of the spectroscopic setup have been described elsewhere.[15] Surface photovoltage was measured on as grown wafers using a Kelvin probe (Besoke Delta Phi Gmbh) in the same spectrometric system under identical illumination conditions as the photocurrent.

Further details of the surface photovoltage method and its physics may be found in Ref. 16.

A photovoltage spectrum measured on an unpatterned area of the wafer, and normalized to the surface transmission, is shown on Fig.

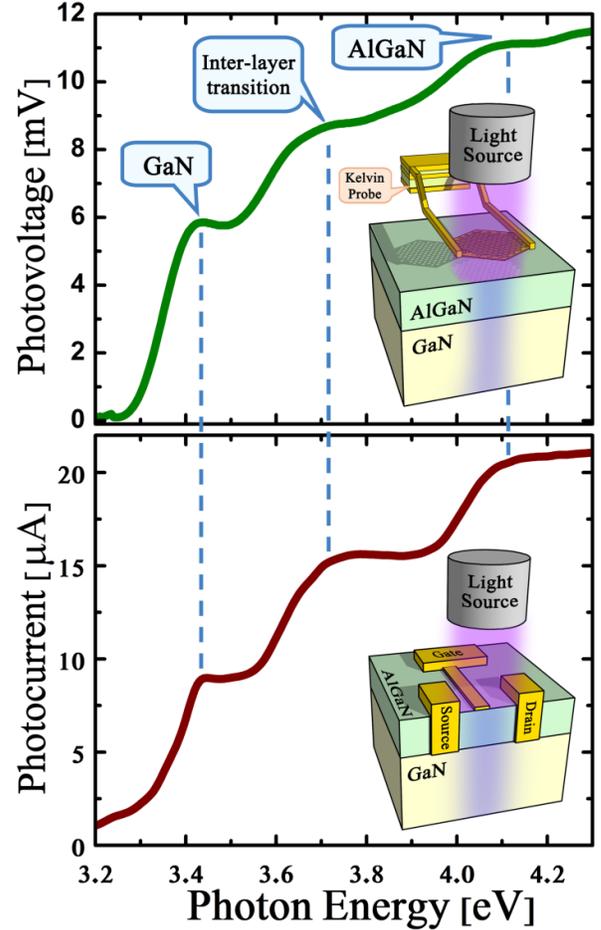

**Fig. 1** <u>Top</u> – surface photovoltage spectrum acquired from an as grown AlGaN/GaN structure grown on sapphire. It shows a set of 3 steps. Each step ends at a transition energy. The first and the last transition energies are the GaN and the AlGaN bandgap, respectively. Between them there supposed to be two inter-layer transitions. However, one of these transitions is weak and only the other can be clearly distinguished (from AlGaN valence band to GaN conduction band). The <u>inset</u> shows the layer structure with a cartoon of the semitransparent Kelvin probe and a spectrometric light source. <u>Bottom</u> – photocurrent measured in the channel of a HEMT transistor fabricated on the same wafer and measured in the same spectroscopic system and under the same conditions as the photovoltage. The photocurrent shows the same steps at the same energies. The <u>inset</u> shows the transistor illuminated from the gate side.

1a. Figure 1b shows a surface-transmission-normalized photocurrent spectrum obtained from a transistor fabricated on the same wafer. Similar step responses of GaN, AlGaN and AlGaN/GaN inter-layer transitions are observed in both spectra. In both spectra the steps are not sharp, but rather preceded by a gradual increase. We have shown before that the slope preceding each photocurrent step is a result of electric-field-assisted absorption, also known as the Franz-Keldysh effect, and may be used to measure the electric field in the corresponding layer.[21] Here, we compare results from the previously established photocurrent method to those obtained from surface photovoltage. To obtain the built-in fields from the photovoltage spectrum, we used Eq. 9 which transforms the slope preceding the bandgap energy into a linear curve. Figure 2 shows the linear form of the photovoltage step response of AlGaN layer (circles). The linearity of the curve provides graphical confirmation to the assumption that the physics underlying the shape of the step is indeed the Franz-Keldysh effect. A solid line shows the linear fit. According to Eq. 9, the linear curve intersects the photon energy axis at the bandgap, while the slope may be used with Eq. 3 to calculate the built-in electric field in the layer. Using this graphical method yields an electric field of *0.596±0.018 MV/cm* in the GaN layer and *1.318±0.031 MV/cm* in the AlGaN layer. For comparison, the electric fields obtained from photocurrent spectrum using the photocurrent model[15] were *0.597±0.036* and *1.374±0.018 MV/cm* for GaN and AlGaN layers, respectively.

Bandgaps obtained from the intersection of linear fit with the photon energy axis for the spectral data shown in Fig. 1 were 4.05 eV for AlGaN and 3.43 eV for GaN. The AlGaN/GaN interlayer transition at 3.68 eV suggests a valence band offset of $\Delta E_V = 0.15\ eV$. The other interlayer transition is rather weak in both spectra. The 2-DEG sheet charge densities calculated using Eq. 10 from the photovoltage and photocurrent data are $9.78(\pm 0.27) \cdot 10^{12}\ cm^{-2}$ and $9.47(\pm 0.32) \cdot 10^{12}\ cm^{-2}$, respectively. The values of the polarization vectors used in the calculation were *-0.0340 C/m²*, *-0.0464 C/m²* and *-0.0098 C/m²* for $P_{SP,GaN}$, $P_{SP,AlGaN}$ and $P_{PE,AlGaN}$ (Al composition of 30%), respectively.[22] The values obtained with the two methods are within measurement error from one another.

Using the photovoltage method on AlGaN/GaN structure, we obtained the bandgaps, band offsets and built-in electric fields in the structure and verified their accuracy comparing them to the same parameters obtained using the photocurrent method. Figure 3 shows the resulting equilibrium band diagram of the HEMT structure that corresponds to these values. The potential in the AlGaN was assumed linear, while for the GaN potential we used a one-dimensional Schrodinger-Poisson solver. The input parameter used in the solver, was the electric field measured in the GaN at the intersection of the first sub-band energy in the GaN quantum well with the conduction band.

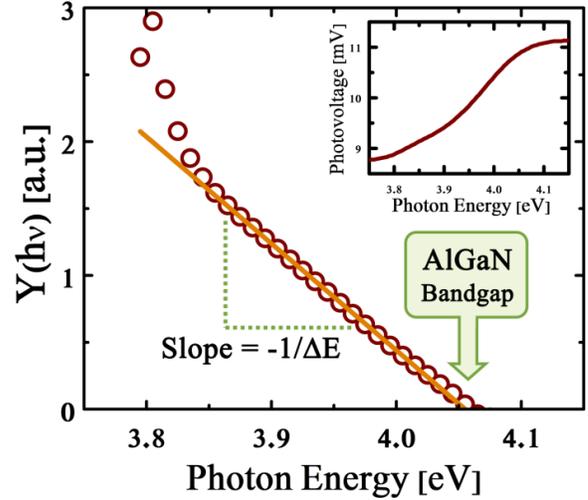

**Fig. 2** A linear representation of the photovoltage step response of AlGaN layer (The original photovoltage response is shown in the inset) using Eq. 9. The intersect of the linear curve with the photon energy axis is at the bandgap of the material. The slope of the linear curve may be used to calculate the built-in field in the layer.

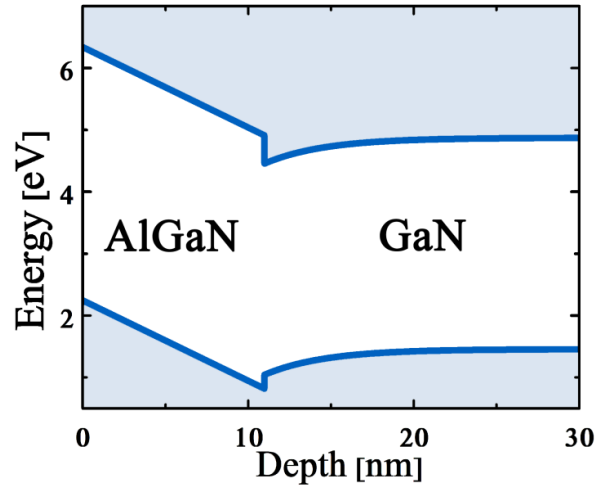

**Fig. 3** An equilibrium band diagram of the AlGaN/GaN HEMT structure calculated from the photovoltage data.

Within our measurement error, both the photocurrent and the photovoltage methods provided the same values of the internal electric fields and the 2-DEG sheet charge density. Although, at first sight, both methods appear to yield very similar spectra, the photovoltage method holds a major advantage over the photocurrent method by being contactless. It does not require the fabrication of a transistor or even of Ohmic contacts. Using photovoltage, the most significant transistor parameters can be measured at the bare wafer stage, before any fabrication step has taken place. Nonetheless, this method is limited to equilibrium conditions. To evaluate the dependence of these parameters on the gate voltage, one would still need the photocurrent method that requires the full fabrication of a transistor. Hence, the photovoltage method may serve for pre-fabrication screening of wafers, whereas the photocurrent method may be used for a full-blown post-fabrications studies. Photovoltage scan of the wafer may provide a map of the wafer to evaluate the lateral uniformity of the 2DEG charge density. Finally, the method presented here is by no means limited to characterization of HEMT structures, but should be suitable to any modern multi-layer device, e.g. heterojuncion bipolar transistor, multi-quantum well LED, etc.

## Acknowledgements

This work was funded by a research grant from the Israeli Ministry of Defence (MAFAT).